\documentclass[conference]{IEEEtran}
\usepackage{cite}

\ifCLASSINFOpdf
  \usepackage[pdftex]{graphicx}
\else
\fi
\ifCLASSOPTIONcompsoc
\else
\fi
\usepackage{stfloats}
\usepackage{url}

\usepackage{booktabs}
\usepackage[frozencache]{minted}

\usepackage{siunitx}
\DeclareSIUnit{\FLOP}{FLOP}
\newcommand{\cwl}[1]{\mintinline{yaml}{#1}}
\usepackage{multicol}
\begin{document}
%
\title{Supercomputing with MPI meets the Common Workflow Language standards: an experience report}

\author{\IEEEauthorblockN{Rupert W. Nash, Nick Brown}
\IEEEauthorblockA{EPCC, The University of Edinburgh\\
Edinburgh, United Kingdom\\
Email: \{r.nash, n.brown\}@epcc.ed.ac.uk}
\and
\IEEEauthorblockN{Michael R. Crusoe}
\IEEEauthorblockA{Vrije Universiteit; Amsterdam, NL \\
ELIXIR-NL/DTL Projects\\
Email: mrc@commonwl.org}
\and
\IEEEauthorblockN{Max Kontak}
\IEEEauthorblockA{DLR German Aerospace Center\\
Institute for Software Technology\\
High-Performance Computing\\
Cologne, Germany}
}

\maketitle

\begin{abstract}
Use of standards-based workflows is still somewhat unusual by high-performance computing users. In this paper we describe the experience of using the Common Workflow Language (CWL) standards to describe the execution, in parallel, of MPI-parallelised applications. In particular, we motivate and describe the simple extension to the specification which was required, as well as our implementation of this within the CWL reference runner. We discuss some of the unexpected benefits, such as simple use of HPC-oriented performance measurement tools, and CWL software requirements interfacing with HPC module systems. We close with a request for comment from the community on how these features could be adopted within versions of the CWL standards.
\end{abstract}


%
\IEEEpeerreviewmaketitle

\section{Introduction}
The use of standards-based workflows in HPC is a small, but growing, area of our community. As developers start to recognise the benefits that workflows can deliver, then they often find that the automation, structure, abstraction, and portability that workflows can provide, delivers significant productivity benefits in the long term. One such user is the VESTEC project, which required the ability to describe the execution of applications, and then invoke them on a supercomputer. The specific requirements in this instance included the ability to describe the inputs, to execute the tool itself, and then to collect the any generated outputs. Furthermore, it was highly desirable that these steps would be represented in a structured, and ideally standard, fashion. VESTEC team members (Nash, Brown, Kontak) found that the \cwl{CommandLineTool} concept from the Common Workflow Language v1.1 standard~\cite{cwl10} and the CWL reference runner ("cwltool")\footnote{\url{https://github.com/common-workflow-language/cwltool}} satisfied these requirements, but with one crucial caveat: namely that many of our tools were MPI-parallelised applications and this is not yet directly supported by the CWL standard.

The VESTEC (Visual Exploration and Sampling Toolkit for Extreme Computing) project aims to fuse HPC with real-time data for urgent decision making for disaster response. This project involves numerous simulation codes to run on HPC machines, and the requirement to couple these together such that the output from one code can be used as an input to others. CWL looked to be an attractive option to provide this in a structured manner on production machines with existing MPI implementations. However a key requirement was the ability to drive this from within a batch submission script, where the batch job itself allocates a number available nodes and starts the CWL tool, and this engine is then able to interact with the MPI runner in order to direct the launch parameters, such as number of MPI processes and placement. Furthermore, this needed to be easily configurable by the user, for each step in the CWL workflow. As it stood, this functionality was not available.

\begin{figure*}[!tb]
    \centering
    \includegraphics[width=\textwidth]{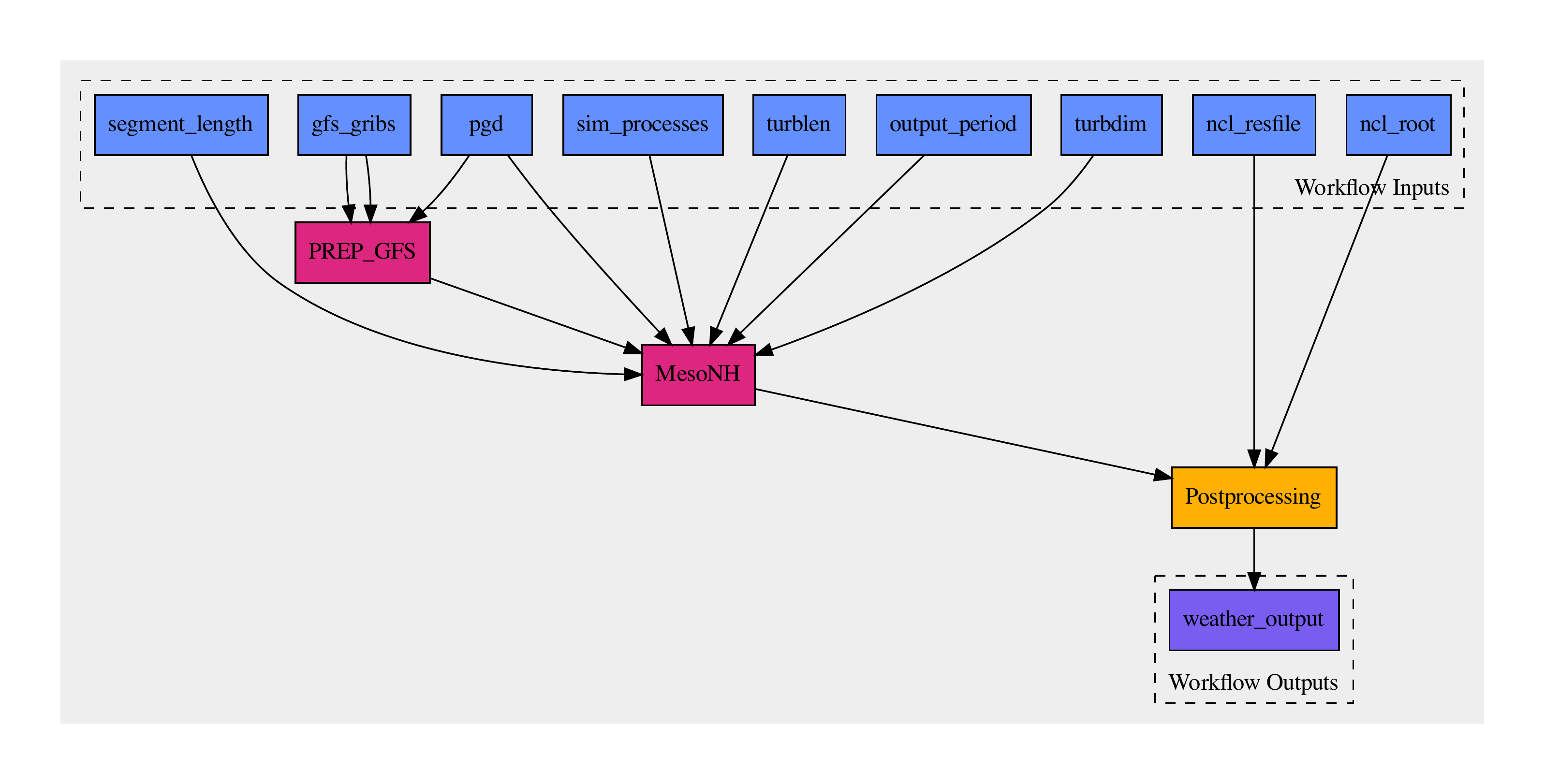}
    \caption{Schematic of a simple workflow used within the VESTEC project. The upper, blue boxes denote inputs, magenta boxes workflow steps which may be run via MPI, yellow boxes conventional workflow steps, and the lower, purple boxes outputs. Arrows show the dependencies within the workflow.}
    \label{fig:mnh-wf}
\end{figure*}

In figure~\ref{fig:mnh-wf} we illustrate a representation of one of the CWL workflows used within the project. This workflow consumes one or more global weather forecasts from the US NOAA Global Forecast System\footnote{\url{https://www.ncdc.noaa.gov/data-access/model-data/model-datasets/global-forcast-system-gfs}}, and interpolates the meterological fields onto the domain for the simulation (specified by the input \verb`pdg`, \emph{i.e.} the physiographic data). 
It then runs the Meso-NH~\cite{mnh54} mesoscale atmospheric simulation application in parallel (as specified by the \verb`sim_processes` input) using the GFS data provided as initial and boundary conditions, for an experiment of simulated duration \verb`segment_length`. The outputs of this simulation are then post-processed by a script into a single netCDF file with the fields of interest for use later in the outer workflow.

The Message Passing Interface (MPI)~\cite{MPI31} is an important standard for parallel programming, especially for large, tightly-coupled simulations; for example a 2017 survey by the US Exascale Computing Project~\cite{ecpsurvey} identified that \emph{all} the responding application development projects under its umbrella were using MPI in some way. This approach works by starting many copies of the same program, which differ only by a unique index (their rank) and providing a mechanism for them to perform point-to-point and collective communication, synchronisation operations, and  IO. Most HPC systems include a low-latency, high-bandwidth interconnect which is not accessed via the usual TCP/IP stack, which the MPI libraries themselves are specifically optimised to take advantage of. 

We have created an extension to the CWL standard to allow a tool description to prepend the platform-specific job launcher and the arguments it requires to the command line of the tool invocation in a way that is orthogonal to the rest of the tool description. This has been implemented within the CWL reference runner, tested on several HPC clusters, and merged into the CWL reference runner behind a feature flag. Furthermore, due to the configurable way in which we have designed the extension, it was trivial to add automatic collection of performance statistics during parallel job execution using performance gathering tools, such as LIKWID\cite{likwid_paper}.

Another challenge to supporting the execution of portable workflows on supercomputers is the requirement for custom-compiled software and the lack of software containers for performance reasons.
This is somewhat orthogonal to the classic CWL approach, as the standards have long supported both software containers and references to the name (and published identifier, if available) of the software tool. The CWL reference runner has a feature which maps these software identifiers to locally available software packages, and loads them in a site-specific way using a local configuration\footnote{\url{https://github.com/common-workflow-language/cwltool/blob/main/README.rst#leveraging-softwarerequirements-beta} using a library from the Galaxy Project\cite{galaxy}}. We have adopted this same approach within the VESTEC system, which ensures that our workflows are portable between target HPC systems.


This paper is organised as follows, in Section \ref{sec:bg} we briefly explored related work and alternatively systems which aim to solve the problem, but often do so in a manner not suited to our requirements in VESTEC. This is followed by Section \ref{sec:imp} which describes the extensions made to the CWL reference runner in order to support a knowledge of MPI and the ability to control important parameters from workflow scripts. Section \ref{sec:performance} then describes the use of these extensions to not only drive the execution of workflow tasks via MPI, but furthermore demonstrates the ability to gather performance metrics by calling into profiling tools, such as LIKWID~\cite{likwid_paper} in this example, which is often an important activity in HPC. Section \ref{sec:conclusions} then concludes this paper, summarising our findings, experience and discussing further work. In particular, we pose questions to the wider HPC workflow community about whether this approach is the correct one for extending the CWL specification.

\section{Related Work}
\label{sec:bg}
Whilst there is an embarrassment of riches\footnote{\url{https://s.apache.org/existing-workflow-systems}} when it comes to selecting a workflow system, many of these were developed with limited single node level concurrency in mind. Therefore, whilst it is often possible to run tasks in parallel as individual tasks which are allocated to multiple cores on the same node, in order to extend to run across modern supercomputers with multiple nodes, more work is required. There have been a number of attempts at this, some more applicable to our specific challenge here than others, but crucially the standardised nature of CWL offers other significant advantages which made it a very attractive technology to adopt.

Workflow systems have followed a number of distinct approaches to supporting execution across multiple nodes. The tightest integrated are those that look to run the different stages in parallel inter-node, with Parsl~\cite{babuji2019parsl} and Swift/T~\cite{armstrong2014compiler} being two examples of this. Swift/T compiles a programmer's Swift code into an MPI program, where the semantics of Swift, which is naturally concurrent, make it ideal as an abstraction around workflows running concurrently. The language schedules the execution of the programmer's statements, as tasks, based upon the availability of input data, and coordinates the distribution and movement of data between the tasks. Implicitly moving data using MPI, the idea is that much of this acts as a wrapper for calls to launch existing tasks.

A major limitation of these tightly coupled approaches is that, by coupling so closely to MPI, some of the limitations of MPI permeate into the workflow system. For instance in Swift/T, the lack of fault tolerance in MPI means that the workflow engine relies upon all tasks completing without error, and has limited recourse if an error occurs inside one of the MPI jobs or the MPI runtime itself. Furthermore, the dynamic job launching capabilities of MPI, which are known to be limited, are not sufficient to fully support the launching of tasks in Swift/T. To address this, the authors proposed an extension to the MPI standard, \mintinline{C}{MPI_Comm_launch}~\cite{wozniak2019mpi}, which allows a child MPI application to be launched inside the resources originally held by processes of a parent MPI application. This has been implemented within a bespoke MPI extension of on a Linux cluster system, which indeed is required for Swift/T, however would not be possible on mainstream production supercomputers that use the PMI interface.

The technologies described above, where the engine translates source code of a programming language into steps is one example of a workflow abstraction. There are other, arguably less tightly coupled, general purpose approaches that support the expression of the stages through configuration files. A number of these, including DARE~\cite{klampanos2019dare} and Pegasus~\cite{deelman2015pegasus}, support execution of their workflows on HPC machines. For instance a specific tool, pegasus-mpi-cluster~\cite{rynge2012enabling}, has been developed for Pegasus which enables it to run high-throughput scientific workflows on HPC systems. MPI is the key enabling technology here, and the engine adopts a scheduler/worker pattern where one of the MPI processes manages the workflow stages to be executed, distributes these amongst the workers, and workers then communicate results back to the scheduler once completed. This is a powerful approach for parallel applications that follow such a pattern of parallelism, however it is not shared by the applications of the VESTEC project. Instead, this project involves codes which predominately follow a geometric decomposition approach, and write their output to the file-system, often in parallel. For such an approach to work, where tasks are mapped to workers, then at a minimum each worker would need to be capable of running over numerous MPI processes within their own communicator, which is not implemented. Furthermore, whilst there are numerous command line options provided by the pegasus-mpi-cluster tool, if users wished to provide further options then this would require in-depth modification of the tool itself.

Pegasus also provides an alternative way of executing on HPC machines via interaction with the batch scheduling system using a number of approaches built atop HTCondor~\cite{thain2005distributed}. Parameters to the MPI job launcher can be provided via the generation of a site catalogue, one for each machine, which specifies the set of applicable configuration options which can be set. This is a nice approach, however rather heavyweight for the fairly simple workflows we have in VESTEC. In fact, we favour following an opposite method, where instead of the workflow system interacting with the batch system to launch specific tasks, it is the execution of the workflow engine itself that is submitted to the batch system. Within the confines of the resources which have been made available, the engine then marshals which MPI processes run and when, directing the configuration of the MPI job launcher based upon the user's workflow description.

Numerous domain specific workflow systems have been developed which, to some extent, support execution over HPC machines. One example of this is the weather and climate community, who make extensive use of workflows to drive their codes. Cycl~\cite{Cylc:2019} is a popular workflow engine in this community, designed specifically with the challenge of cyclical workflows in mind. Used in production by organisations which include the UK Met Office and NIWA, the technology is also able to manage the submission of jobs via common HPC batch systems, however it has no specific knowledge of the underlying MPI \emph{per se}. Therefore, finer-grained control around the behaviour of MPI between allocated nodes, for instance the ratio of MPI processes to threads, is not possible, which is required by VESTEC.

ecFlow~\cite{bahra2011managing} has been developed by ECMWF and in this technology the launching is undertaken by shell scripts, which themselves are called by ecFlow. It is therefore these scripts, rather than the workflow system itself, that has knowledge of how the code should be launched, which the workflow engine has no visibility of. Furthermore, elements of ecFlow must be installed on the target system to enable task communications. Autosubmit~\cite{manubens2016seamless} is another domain specific workflow technology developed for weather and climate. Delivered as a lightweight Python module, unlike Cycl or ecFlow, the target HPC system need have no specific support installed for the workflow engine. Built on top of Simple API for Grid Applications (SAGA)~\cite{goodale2006saga}, 

A number of HPC codes have recognised the benefits of workflows, but developed an approach which is very application specific. One example of this is Gromacs~\cite{javanainen2017atomistic}, a popular molecular dynamics application which is highly parallelised and runs well on  many thousands of CPU cores. The BioExcel Building Blocks (BioBB) library~\cite{andrio2019bioexcel} developed a workflow approach with the aim of bringing biomolecular simulations closer to the bioinformatics way of working. Scientists build individual blocks by wrapping software components in Python code, with an overarching workflow system such as Galaxy, PyCOMPSs, and CWL then driving these. However, it is these wrapper scripts that actually handle the job launching with MPI and the associated parameters, with the workflow system itself has no visibility of this.

The Common Workflow Language open standards\cite{cwl10} are a set of community driven specifications for describing command line tools as typed functions and the workflows made from these tools. While CWL originated in the bioinformatics community, it was an early goal to not make the standards specific to bioinformatics or even the life sciences. There are many implementations\footnote{\url{https://www.commonwl.org/#Implementations}} of the CWL standards with a variety of licenses: proprietary software as a service (products from Seven Bridges Genomics and others), commercially supported open source (Arvados\footnote{\url{https://arvados.org/}}), and academic open source. The compliance of these projects to the CWL standards can be tested using the CWL conformance tests\footnote{\url{https://github.com/common-workflow-language/common-workflow-language/blob/main/CONFORMANCE_TESTS.md}}. The Toil\cite{toil} project has a "toil-cwl-runner" workflow executor that passes all of the CWL v1.0 and most of CWL v1.1 and CWL v1.2 conformance tests; it has backend support for many HPC/HTC job scheduling systems and can also manage an ephemeral cluster on commercial cloud computing systems. The CWL reference runner "cwltool" aims at providing a test bed for implementation of new CWL features, but does not aim to be a complete production level workflow management platform. For example, the CWL reference runner does not interface with job schedulers or do any type of remote execution.

\section{Design and Implementation}
\label{sec:imp}
It is worth noting that while the MPI standard deliberately does not require a particular method for starting MPI programs, they are typically started with a job launcher command similar to that recommended in the standard: \texttt{mpiexec -n <num processes> <executable> <program arguments>}. Additionally, the job launcher may not run on one of the machine that will actually execute the program (indeed this is the case for one of the machines used by VESTEC: ARCHER, a Cray XC30 system). Further, the job launch command may require environment variables to be set in order to communicate, \emph{e.g.} which nodes are available for use, as is common on clusters using SLURM.

Since CWL lacks structured MPI support, the so-called base command of the tool description would become either: the platform specific \verb`mpiexec`, thus negating portability of the tool description and relegating the actual application to merely an argument; or a custom wrapper script which interposes between CWL and the tool imposing an greater burden upon either the tool or tool description author.

Before extending the CWL specification and reference runner, we first attempted to use the CWL feature of using JavaScript embedded in the tool description to programmatically insert the necessary MPI job launch commands to the front of the command line string. Unfortunately this required a somewhat convoluted job description file, as can be seen by comparing examples (a) and (b) from listing~\ref{list:hello} which show, respectively, the (serial) "Hello world" example from the CWL tutorial\footnote{\url{https://www.commonwl.org/user_guide/02-1st-example/index.html}} and a version which launches multiple copies of the same in parallel via MPI.

\begin{listing*}[tb]
\begin{multicols*}{3}
\small\centering
\begin{minted}{yaml}
cwlVersion: v1.0
class: CommandLineTool







inputs:
  message:
    type: string
    inputBinding:
      position: 1



baseCommand: echo


outputs: []
\end{minted}
(a)
\columnbreak

\begin{minted}{yaml}
cwlVersion: v1.0
class: CommandLineTool
requirements:
  InlineJavascriptRequirement:
    expressionLib:
      - $include: mpi.js
  SchemaDefRequirement:
    types:
      - $import: mpi.yml
inputs:
  message:
    type: string
    inputBinding:
      position: 1
  mpi:
    type: mpi.yml#mpiInfo
    default: {}
arguments:
  - position: 0
    valueFrom: $(mpi.run("echo"))
outputs: []
\end{minted}
(b)
\columnbreak

\begin{minted}{yaml}
cwlVersion: v1.0
class: CommandLineTool
$namespaces:
  cwltool: http://commonwl.org/cwltool#

requirements:
  cwltool:MPIRequirement:
    processes: $(inputs.nproc)

inputs:
  message:
    type: string
    inputBinding:
      position: 1
  nproc:
    type: int

baseCommand: echo


outputs: []
\end{minted}
(c)
\end{multicols*}
\caption{The first, "Hello World" example from the CWL tutorial in three ways. (a) original, serial version; (b) our interim, MPI parallel implementation; (c) final version using \cwl{MPIRequirement}. In each case, the tool accepts one argument which is echoed to standard output. In cases (b) and (c), this is run in parallel via the MPI launcher.}\label{list:hello}
\end{listing*}

This approach has a number of further drawbacks, such as requiring that a JavaScript engine be available, not allowing unknown (to the tool description author) environment variables to be set, and interacting poorly with software container runtimes. After engaging with the CWL community and the CWL Project Lead (Crusoe), it was clear that the job runner needed to understand that this was different from normal execution, in much that same way that execution within a container was different.

We formulated the following requirements on this extension to the specification. That tool descriptions:
\begin{itemize}
    \item must opt in to potentially being run via MPI;
    \item must allow for this to be disabled;
    \item must be able to control number of processes either directly or via an input;
    \item must remain the same for different execution machines;
    \item should be as close to a non-MPI version of the same tool as practical.
\end{itemize}
And also that the runner also needs to provide a configuration mechanism:
\begin{itemize}
    \item to specify the platform specific launcher;
    \item to specify how to set the number of processes;
    \item to add any further flags required;
    \item to pass through or set any environment variables required.
\end{itemize}

In a more general case, for example of hybrid MPI + OpenMP parallelism, correctly launching an application requires knowledge of multiple things. First, the execution hardware: what is the memory architecture of a node (number of cores per NUMA region, how many NUMA regions per socket, number of sockets per node) and their number? Second, how does one set the number of MPI processes (typically via command line argument) and OpenMP threads (typically via the \verb`OMP_NUM_THREADS` environment variable)? Third, how does one specify process and thread binding rules? Fourth, what application specific knowledge must be considered? A similar exercise would have to completed for GPU accelerated applications. Because of the large amount complexity involved in a general solution and the fact that our use case typically requires only one parallel execution per workflow (allowing us to use the extra configuration options to supply this), we simply defer this to future work.

CWL supports the concept of a requirement which "modifies the semantics or runtime environment of a process"\footnote{\url{https://www.commonwl.org/v1.1/CommandLineTool.html#Requirements_and_hints}}, which is a natural fit for the task our problem. The minimum features we need are to enable the requirement and to pass through the number of MPI processes to start (we treat the case of zero processes requested as being equivalent to disabling the requirement). The number of processes can either be a plain integer or a CWL \cwl{Expression} which evaluates to an integer.

The formal definition of CWL is given in the SALAD~\cite{salad} schema definition language, which precisely defines the  keys and their types. In listing~\ref{list:mpireq} we show a simplified\footnote{The full specification for MPIRequirement is at \url{https://github.com/common-workflow-language/cwltool/blob/83038feb2a6fc3bab952e1ecc2a11bfbc8c557b4/cwltool/extensions-v1.1.yml#L48}} version of the addition we made to the CWL specification. We use the SALAD feature of inheritance to declare that this type can be used in the \cwl{requirements} or \cwl{hints} section of a tool description. We then declare that this object has two fields: \cwl{class} which is always the string \cwl{"MPIRequirement"}; and \cwl{processes} which is either an integer or a string. We use a simple string instead of an \cwl{Expression} due to a bug\footnote{\url{https://github.com/common-workflow-language/schema_salad/issues/326}} in the library which implements the SALAD specification. Setting the attribute \cwl{inVocab} to a false declares that this object should not be added to the standard CWL specification but remain in the namespace of the reference runner (in this case \url{http://commonwl.org/cwltool}).

\begin{listing}
\begin{minted}{yaml}
- name: MPIRequirement
  type: record
  extends: cwl:ProcessRequirement
  inVocab: false
  fields:
    - name: class
      type: string
      jsonldPredicate:
        "_id": "@type"
        "_type": "@vocab"
    - name: processes
      type: [int, string]
\end{minted}
\caption{SALAD-YAML description of the \cwl{MPIRequirement}. Note that labels and documentation strings have been removed for clarity.}
\label{list:mpireq}
\end{listing}

A simple example tool description which uses this extension is shown in listing~\ref{list:hello}(c). In this case, the user must provide an input giving the number of MPI processes to start. The MPI requirement object then uses this input to evaluate the expression given as the value of \cwl{processes}.

The cwltool reference runner is open source (Apache 2 licensed) and written in Python (for versions $\geq$ 3.5). The runner is written in a clear, object-orientated style, with an extensive unit-test suite, making development relatively easy. We added a command line option (\mintinline{sh}{--mpi-config-file}) to cwltool to accept a simple YAML file containing the platform configuration data. We show descriptions of the allowed keys, their types, and their default values if omitted in table~\ref{tab:conf-keys}. 

\begin{table*}[tb]
\centering
\caption{Description of allowed keys in the MPI platform configuration file}\label{tab:conf-keys}
\begin{tabular}{lllp{0.5\textwidth}}
\toprule
Key & Type & Default & Description\\
\midrule
\mintinline{python}{runner} & \mintinline{python}{str} & \mintinline{python}{"mpirun"} & The primary command to use. \\
\mintinline{python}{nproc_flag} & \mintinline{python}{str} & \mintinline{python}{"-n"} & Flag to set number of processes to start. \\
\mintinline{python}{default_nproc} & \mintinline{python}{int} & \mintinline{python}{1} & Default number of processes.\\
\mintinline{python}{extra_flags} & \mintinline{python}{List[str]} & \mintinline{python}{[]} & A list of any other flags to be added to the runner's command line before the \cwl{baseCommand}.\\
\mintinline{python}{env_pass} & \mintinline{python}{List[str]} & \mintinline{python}{[]} & A list of environment variables that should be passed from the host environment through to the tool (\emph{e.g.} giving the nodelist as set by your scheduler). \\
\mintinline{python}{env_pass_regex} & \mintinline{python}{List[str]} & \mintinline{python}{[]} & A list of Python regular expressions that will be matched against the host's environment. Those that match will be passed through.\\
\mintinline{python}{env_set} & \mintinline{python}{Mapping[str,str]} & \mintinline{python}{{}} & A dictionary whose keys are the environment variables set and the values being the values.\\
\bottomrule
\end{tabular}
\end{table*}

Within the runner, this argument, if present, is used to configure the MPI runtime. When a tool is actually executed, the runner checks for the \cwl{MPIRequirement}, evaluates the processes attribute and if present and non-zero, it uses the configuration data, as described in table~\ref{tab:conf-keys}, to prepend the appropriate strings to the front of the command line and alter the runtime environment. 

In keeping with the coding standards for cwltool, we implemented fourteen unit tests, which are run in a continuous integration system. Since this lacks an MPI library, we also had to provide a mock MPI job launcher. Our work has been merged\footnote{\url{https://github.com/common-workflow-language/cwltool/pull/1276}} into the main branch of the repository and released. We performed some simple tests of running containerised parallel applications: we could not make this work using Docker, however we had success in simple cases using Singularity~\cite{singularity}.

We also tested the extension by applying it within the VESTEC project, to describe the individual tasks that are performed. In the initial use-case evaluated this was an urgent simulation of a forest fire using real time weather forecast data, the workflow for which is shown above in figure~\ref{fig:mnh-wf}. We could use a single platform-dependent configuration file to run multiple MPI programs (steps \verb`PREP_GFS` and \verb`MesoNH`) with different numbers of processes.

One feature we have yet to implement is allowing users to specify tool-dependent overrides to the platform configuration data. For example one might wish to create a workflow with one pure-MPI code and another that uses multiple threads per MPI-process; these would likely require to be launched with different flags, for example to change the number of processes per node and alter the thread binding, which is not currently possible.  We note that nevertheless individual steps within a workflow can still start different numbers of MPI processes; what we lack is a fine-grained method for control of more advanced, often machine specific MPI options.

\section{MPI-aware Performance monitoring}
\label{sec:performance}

Monitoring the performance of parallel applications is particularly important when they may be running across thousands of cores and is very common in the HPC and supercomputing community.
In a scenario where a workflow may execute very many parallel jobs this is even more so.

There are several tools that can be used to measure the performance of applications, for example, the open-source Linux kernel tool \texttt{perf}~\cite{perf} or the proprietary Intel VTune~\cite{vtune}.
For an integration of performance measurements with the CWL extension presented in this paper, we have decided to use the open-source LIKWID tool suite~\cite{likwid_paper, likwid_web}, which contains a set of command-line tools that enable us to read out performance counters of the CPU easily. 
For example, it provides a drop-in replacement for \texttt{mpirun} called \texttt{likwid-mpirun}, which can be provided directly in the CWL MPI configuration YAML file as the platform-specific runner.

Moreover, in the VESTEC project, where this functionality will be used, we want to be able to automatically process performance measurement data.
LIKWID allows for this by providing so-called output filters, which generate output in machine-readable formats (e.\,g., JSON) instead of human-readable command-line output.
This enables us to automatically process the created data.
Unfortunately, \texttt{likwid-mpirun} does not yet support output filters.
However, one can still combine the platform specific runner with the more basic \texttt{likwid-perfctr} tool, which can measure the performance for a single MPI rank, such that one JSON output file with performance data is created for each rank, which can then be combined to get the data for all MPI ranks.
Our implementation of \cwl{MPIRequirement} does also easily allows us to configure this combination of different tools as can be seen in Listing~\ref{list:mpiconf_likwid}.

\begin{listing}[b]
\begin{minted}{yaml}
runner: srun
extra_flags: [
  "likwid-perfctr", 
  "-C", "L:N:0", 
  "-g", "FLOPS_DP", 
  "-o", "/output/path/likwid_%j_%h_%r.json"
  ]
nproc_flag: -n
env_pass_regex: ["SLURM_.*"]
\end{minted}
\caption{The MPI configuration file used for the execution of an MPI application with CWL on a cluster that uses SLURM as batch scheduler (therefore, the runner is \texttt{srun}) and providing machine-readable performance data files through the \texttt{likwid-perfctr} tool. The so-called performance group \texttt{FLOPS\_DP} has been chosen, such that LIKWID reports double precision floating point performance.}
\label{list:mpiconf_likwid}
\end{listing}

As an example, we have created a CWL file for the HPCG benchmark~\cite{HPCG}, which also allows for the configuration of the problem size and target run time parameters. This was chosen since the application reports its own estimates of the floating point performance achieved, allowing us to have some confidence that the reported numbers are correct.
We have executed the benchmark in an MPI-only manner on one and two nodes of a cluster at German Aerospace Center, where each node consists of 4 Intel Xeon Gold 6132 CPUs with 14 cores each, yielding a total of 56 and 112 cores, respectively.
As the MPI configuration file, we have used the one already presented in Listing~\ref{list:mpiconf_likwid}.
After running the benchmark, LIKWID provides us with the data about double precision performance, shown in Table~\ref{tab:results}.

\begin{table*}[tb]
    \centering
    \caption{Results reported by LIKWID for the HPCG benchmark when executed with CWL on one and two nodes of the used cluster.
    The table shows the total number of cores, the total number of floating point operations per second, the mean and standard deviation per rank as well as the total number of scalar and vectorized micro-operations.}
    \begin{tabular}{rrrrrrr}\toprule
    & HPCG reported & \multicolumn{5}{c}{LIWKID reported} \\
    & Performance / \si{\giga\FLOP\per\second} & \multicolumn{3}{c}{Performance / \si{\giga\FLOP\per\second}} & \multicolumn{2}{c}{Micro operation rate / \si{\giga\second^{-1}}} \\
    Cores &  & Total & Rank mean & Rank s.d. & Total scalar & Total vector \\\midrule
    56 & 38.0 & 39.7 & 0.71 & 0.002 & 39.0 & 0.31 \\
    112 & 71.7 & 74.6 & 0.67 & 0.002 & 73.5 & 0.58 \\\bottomrule
    \end{tabular}
    \label{tab:results}
\end{table*}

LIKWID reports a total performance of \SI{74.6}{\giga\FLOP\per\second} on two nodes, whereas the HPCG benchmark itself reports \SI{71.7}{\giga\FLOP\per\second} (based on its measured run time and estimate of the number of operations performed), which can be attributed to measurement errors and the fact that \texttt{likwid-perfctr} acts as a wrapper for the whole HPCG binary including computation of the performance statistics, set up times etc., which might yield more floating point operations.
The same is true for the results gathered when running the benchmark on a single node.
Furthermore, from the collected data we can observe that the HPCG benchmark code has very good load balancing among the cores, since the standard deviation of the performance is very small in both cases.
We can also see that the number of floating-point operations per rank only drops by a small amount when using two instead of one node, which gives a hint to a good strong scaling of the benchmark code.
Of course this needs to analysed in more detail for a reliable assessment of the scaling properties.
In Table~\ref{tab:results}, we have also provided the measured number of scalar and packed micro-operations, which indicate that the compiler was not able to generate SIMD (single instruction, multiple data) instructions from the HPCG code efficiently. This gives a hint to possible performance optimizations.

All in all, in this section we have shown a simple proof-of-concept on how the presented CWL extension can be used for an MPI-parallel application and how easily the flexible design of the extension allows us to perform performance measurements.

\section{Conclusion}
\label{sec:conclusions}

In this paper we have presented our experience of working between two quite disjoint communities, HPC and workflows. The VESTEC authors (Nash, Brown, and Kontak) could reasonably be defined as "traditional HPC" researchers and their experience is in writing, understanding, optimising and using parallel applications on supercomputers. Whereas Crusoe's experience however comes from bioinformatics and scientific workflow communities where the need for (and the use of) standards-based approaches is more significant.

The VESTEC authors were pleasantly surprised to find that these tools were powerful and flexible enough to be altered for the purposes described above. We serendipitously found that the software requirements feature of CWL could help make our tool descriptions more portable, and that use of multi-step CWL workflows could simplify our larger workflow system by combining pre- and post-processing steps within a single invocation of a CWL runner. The CWL community is pleased that their efforts to enable the CWL standards to be useful beyond the original bioinformatics users is yet again proven worth while.

Together, we have developed an extension for the Common Workflow Language to specify that a tool requires the use of an MPI job launcher and implemented this within the CWL reference runner. While we acknowledge that this is not as of yet feature complete, it does allow interested users to configure a complex, parallel application while not imposing a large burden on the implementation of the CWL specification, as demonstrated by our concise implementation which has been released within the reference runner.

Therefore we believe that the work described in this paper is, as it stands, of use to the HPC community driving their codes via workflows. The ultimate aim is for this to become part of the CWL specification, and to this end we will extend the MPI platform configuration format to allow overriding for different steps, likely using the reference implementation's overrides extension as inspiration. This will allow tool description authors and users to combine parallel applications which require different launch options within a single workflow. Two further straightforward features we plan to add are: first, to allow specification of the revision of MPI standard required by a tool; and, second, to allow specification of the level of thread support required, as MPI applications can require one of four distinct levels.

Before acceptance to the standard, we believe it important to explore how best to implement the use of \cwl{MPIRequirement} simultaneously with software container engines that include Docker, Singularity, udocker, runc, and others. Likewise we plan to test in one or more cloud environments that offer managed MPI services. Bother exercises will improve the specification and implementation of the \cwl{MPIRequirement} and provide guidance for others who wish to implement the extension.

To further demonstrate the applicability of such a \cwl{MPIRequirement} language extension, another next step is to engage with other implementations of the CWL specification. This will enable the production of one or more independent, proof-of-concept, implementations of our extension.

With our primary goal being to submit the \cwl{MPIRequirement} (or some equivalent) feature for inclusion into a future version of the CWL standard, we aim for this paper to serve, in part, as a request for comment from the wider HPC and workflow communities. We believe that the work explored here has a real world application to further exploiting HPC machines via workflows, and as such the authors would be grateful to receive any comments or suggestions, either via email or public CWL forums\footnote{\url{https://www.commonwl.org/#Support}}.

\section*{Acknowledgment}
This work was funded under the EU FET VESTEC H2020 project, grant agreement number 800904.
This work used the ARCHER UK National Supercomputing Service (\url{http://www.archer.ac.uk}).

\IEEEtriggeratref{9}


\bibliographystyle{IEEEtran}
\bibliography{refs}

\begin{thebibliography}{10}
\providecommand{\url}[1]{#1}
\csname url@samestyle\endcsname
\providecommand{\newblock}{\relax}
\providecommand{\bibinfo}[2]{#2}
\providecommand{\BIBentrySTDinterwordspacing}{\spaceskip=0pt\relax}
\providecommand{\BIBentryALTinterwordstretchfactor}{4}
\providecommand{\BIBentryALTinterwordspacing}{\spaceskip=\fontdimen2\font plus
\BIBentryALTinterwordstretchfactor\fontdimen3\font minus
  \fontdimen4\font\relax}
\providecommand{\BIBforeignlanguage}[2]{{%
\expandafter\ifx\csname l@#1\endcsname\relax
\typeout{** WARNING: IEEEtran.bst: No hyphenation pattern has been}%
\typeout{** loaded for the language `#1'. Using the pattern for}%
\typeout{** the default language instead.}%
\else
\language=\csname l@#1\endcsname
\fi
#2}}
\providecommand{\BIBdecl}{\relax}
\BIBdecl

\bibitem{cwl10}
\BIBentryALTinterwordspacing
P.~Amstutz, M.~R. Crusoe, N.~Tijanić, B.~Chapman, J.~Chilton, M.~Heuer,
  A.~Kartashov, D.~Leehr, H.~Ménager, M.~Nedeljkovich, M.~Scales,
  S.~Soiland-Reyes, and L.~Stojanovic, ``Common workflow language, v1.0.''
  [Online]. Available: \url{https://w3id.org/cwl/v1.0/}
\BIBentrySTDinterwordspacing

\bibitem{mnh54}
\BIBentryALTinterwordspacing
C.~Lac, J.-P. Chaboureau, V.~Masson, J.-P. Pinty, P.~Tulet, J.~Escobar,
  M.~Leriche, C.~Barthe, B.~Aouizerats, C.~Augros, P.~Aumond, F.~Auguste,
  P.~Bechtold, S.~Berthet, S.~Bielli, F.~Bosseur, O.~Caumont, J.-M. Cohard,
  J.~Colin, F.~Couvreux, J.~Cuxart, G.~Delautier, T.~Dauhut, V.~Ducrocq, J.-B.
  Filippi, D.~Gazen, O.~Geoffroy, F.~Gheusi, R.~Honnert, J.-P. Lafore,
  C.~Lebeaupin~Brossier, Q.~Libois, T.~Lunet, C.~Mari, T.~Maric, P.~Mascart,
  M.~Mog\'e, G.~Molini\'e, O.~Nuissier, F.~Pantillon, P.~Peyrill\'e,
  J.~Pergaud, E.~Perraud, J.~Pianezze, J.-L. Redelsperger, D.~Ricard,
  E.~Richard, S.~Riette, Q.~Rodier, R.~Schoetter, L.~Seyfried, J.~Stein,
  K.~Suhre, M.~Taufour, O.~Thouron, S.~Turner, A.~Verrelle, B.~Vi\'e,
  F.~Visentin, V.~Vionnet, and P.~Wautelet, ``Overview of the {Meso-NH} model
  version 5.4 and its applications,'' \emph{Geoscientific Model Development},
  vol.~11, no.~5, pp. 1929--1969, 2018. [Online]. Available:
  \url{https://gmd.copernicus.org/articles/11/1929/2018/}
\BIBentrySTDinterwordspacing

\bibitem{MPI31}
``{MPI}: A message-passing interface standard version 3.1,'' {MPI Forum}, Tech.
  Rep., 2015.

\bibitem{ecpsurvey}
\BIBentryALTinterwordspacing
D.~E. Bernholdt, S.~Boehm, G.~Bosilca, M.~Gorentla~Venkata, R.~E. Grant,
  T.~Naughton, H.~P. Pritchard, M.~Schulz, and G.~R. Vallee, ``A survey of
  {MPI} usage in the {US} exascale computing project,'' \emph{Concurrency and
  Computation: Practice and Experience}, vol.~32, no.~3, p. e4851, 2020, e4851
  cpe.4851. [Online]. Available:
  \url{https://onlinelibrary.wiley.com/doi/abs/10.1002/cpe.4851}
\BIBentrySTDinterwordspacing

\bibitem{likwid_paper}
J.~Treibig, G.~Hager, and G.~Wellein, ``{LIKWID}: A lightweight
  performance-oriented tool suite for x86 multicore environments,'' in
  \emph{Proceedings of PSTI2010, the First International Workshop on Parallel
  Software Tools and Tool Infrastructures}, San Diego CA, 2010.

\bibitem{galaxy}
\BIBentryALTinterwordspacing
E.~Afgan, D.~Baker, B.~Batut, M.~van~den Beek, D.~Bouvier, M.~\v{C}ech,
  J.~Chilton, D.~Clements, N.~Coraor, B.~A. Gr\"uning, A.~Guerler,
  J.~Hillman-Jackson, S.~Hiltemann, V.~Jalili, H.~Rasche, N.~Soranzo,
  J.~Goecks, J.~Taylor, A.~Nekrutenko, and D.~Blankenberg, ``The {Galaxy}
  platform for accessible, reproducible and collaborative biomedical analyses:
  2018 update,'' \emph{Nucleic Acids Res.}, vol.~46, no.~W1, pp. W537--W544,
  2018. [Online]. Available: \url{https://doi.org/10.1093/nar/gky379}
\BIBentrySTDinterwordspacing

\bibitem{babuji2019parsl}
Y.~Babuji, A.~Woodard, Z.~Li, D.~S. Katz, B.~Clifford, R.~Kumar, L.~Lacinski,
  R.~Chard, J.~M. Wozniak, I.~Foster \emph{et~al.}, ``Parsl: Pervasive parallel
  programming in {P}ython,'' in \emph{Proceedings of the 28th International
  Symposium on High-Performance Parallel and Distributed Computing}, 2019, pp.
  25--36.

\bibitem{armstrong2014compiler}
T.~G. Armstrong, J.~M. Wozniak, M.~Wilde, and I.~T. Foster, ``Compiler
  techniques for massively scalable implicit task parallelism,'' in
  \emph{SC'14: Proceedings of the International Conference for High Performance
  Computing, Networking, Storage and Analysis}.\hskip 1em plus 0.5em minus
  0.4em\relax IEEE, 2014, pp. 299--310.

\bibitem{wozniak2019mpi}
J.~M. Wozniak, M.~Dorier, R.~Ross, T.~Shu, T.~Kurc, L.~Tang, N.~Podhorszki, and
  M.~Wolf, ``{MPI} jobs within {MPI} jobs: A practical way of enabling
  task-level fault-tolerance in {HPC} workflows,'' \emph{Future Generation
  Computer Systems}, vol. 101, pp. 576--589, 2019.

\bibitem{klampanos2019dare}
I.~Klampanos, A.~Davvetas, A.~Gem{\"u}nd, M.~Atkinson, A.~Koukourikos,
  R.~Filgueira, A.~Krause, A.~Spinuso, A.~Charalambidis, F.~Magnoni
  \emph{et~al.}, ``Dare: A reflective platform designed to enable agile
  data-driven research on the cloud,'' in \emph{2019 15th International
  Conference on eScience (eScience)}.\hskip 1em plus 0.5em minus 0.4em\relax
  IEEE, 2019, pp. 578--585.

\bibitem{deelman2015pegasus}
E.~Deelman, K.~Vahi, G.~Juve, M.~Rynge, S.~Callaghan, P.~J. Maechling,
  R.~Mayani, W.~Chen, R.~F. Da~Silva, M.~Livny \emph{et~al.}, ``Pegasus, a
  workflow management system for science automation,'' \emph{Future Generation
  Computer Systems}, vol.~46, pp. 17--35, 2015.

\bibitem{rynge2012enabling}
M.~Rynge, S.~Callaghan, E.~Deelman, G.~Juve, G.~Mehta, K.~Vahi, and P.~J.
  Maechling, ``Enabling large-scale scientific workflows on petascale resources
  using mpi master/worker,'' in \emph{Proceedings of the 1st Conference of the
  Extreme Science and Engineering Discovery Environment: Bridging from the
  eXtreme to the campus and beyond}, 2012, pp. 1--8.

\bibitem{thain2005distributed}
D.~Thain, T.~Tannenbaum, and M.~Livny, ``Distributed computing in practice: the
  condor experience,'' \emph{Concurrency and computation: practice and
  experience}, vol.~17, no. 2-4, pp. 323--356, 2005.

\bibitem{Cylc:2019}
H.~{Oliver}, M.~{Shin}, D.~{Matthews}, O.~{Sanders}, S.~{Bartholomew},
  A.~{Clark}, B.~{Fitzpatrick}, R.~{van Haren}, R.~{Hut}, and N.~{Drost},
  ``Workflow automation for cycling systems,'' \emph{Computing in Science
  Engineering}, vol.~21, no.~4, pp. 7--21, 2019.

\bibitem{bahra2011managing}
A.~Bahra, ``Managing work flows with {ecFlow},'' \emph{ECMWF Newsl}, vol. 129,
  pp. 30--32, 2011.

\bibitem{manubens2016seamless}
D.~Manubens-Gil, J.~Vegas-Regidor, C.~Prodhomme, O.~Mula-Valls, and F.~J.
  Doblas-Reyes, ``Seamless management of ensemble climate prediction
  experiments on {HPC} platforms,'' in \emph{2016 International Conference on
  High Performance Computing \& Simulation (HPCS)}.\hskip 1em plus 0.5em minus
  0.4em\relax IEEE, 2016, pp. 895--900.

\bibitem{goodale2006saga}
T.~Goodale, S.~Jha, H.~Kaiser, T.~Kielmann, P.~Kleijer, G.~Von~Laszewski,
  C.~Lee, A.~Merzky, H.~Rajic, and J.~Shalf, ``{SAGA}: A simple {API} for grid
  applications. high-level application programming on the grid,''
  \emph{Computational Methods in Science and Technology}, vol.~12, no.~1, pp.
  7--20, 2006.

\bibitem{javanainen2017atomistic}
M.~Javanainen, I.~Vattulainen, and L.~Monticelli, ``On atomistic models for
  molecular oxygen,'' \emph{The Journal of Physical Chemistry B}, vol. 121,
  no.~3, pp. 518--528, 2017.

\bibitem{andrio2019bioexcel}
P.~Andrio, A.~Hospital, J.~Conejero, L.~Jord{\'a}, M.~Del~Pino, L.~Codo,
  S.~Soiland-Reyes, C.~Goble, D.~Lezzi, R.~M. Badia \emph{et~al.}, ``{BioExcel}
  {B}uilding {B}locks, a software library for interoperable biomolecular
  simulation workflows,'' \emph{Scientific data}, vol.~6, no.~1, pp. 1--8,
  2019.

\bibitem{toil}
J.~Vivian, A.~A. Rao, F.~A. Nothaft, C.~Ketchum, J.~Armstrong, A.~Novak,
  J.~Pfeil, J.~Narkizian, A.~D. Deran, A.~Musselman-Brown, H.~Schmidt,
  P.~Amstutz, B.~Craft, M.~Goldman, K.~Rosenbloom, M.~Cline, B.~O'Connor,
  M.~Hanna, C.~Birger, W.~J. Kent, D.~A. Patterson, A.~D. Joseph, J.~Zhu,
  S.~Zaranek, G.~Getz, D.~Haussler, and B.~Paten, ``Toil enables reproducible,
  open source, big biomedical data analyses,'' \emph{Nature Biotechnology},
  vol.~35, no.~4, pp. 314--316, 2017.

\bibitem{salad}
\BIBentryALTinterwordspacing
P.~Amstutz, ``Semantic annotations for linked {A}vro data ({SALAD}).''
  [Online]. Available: \url{https://www.commonwl.org/v1.1/SchemaSalad.html}
\BIBentrySTDinterwordspacing

\bibitem{singularity}
\BIBentryALTinterwordspacing
G.~M. Kurtzer, V.~Sochat, and M.~W. Bauer, ``Singularity: Scientific containers
  for mobility of compute,'' \emph{PLOS ONE}, vol.~12, no.~5, pp. 1--20, 05
  2017. [Online]. Available: \url{https://doi.org/10.1371/journal.pone.0177459}
\BIBentrySTDinterwordspacing

\bibitem{perf}
\BIBentryALTinterwordspacing
{perf: {L}inux profiling with performance counters}. Accessed 2020-08-25.
  [Online]. Available: \url{https://perf.wiki.kernel.org/index.php/Main\_Page}
\BIBentrySTDinterwordspacing

\bibitem{vtune}
\BIBentryALTinterwordspacing
{Intel VTune Profiler}. Accessed 2020-08-25. [Online]. Available:
  \url{https://software.intel.com/content/www/us/en/develop/tools/vtune-profiler.html}
\BIBentrySTDinterwordspacing

\bibitem{likwid_web}
\BIBentryALTinterwordspacing
{LIKWID Performance Tools}. Accessed 2020-08-25. [Online]. Available:
  \url{https://hpc.fau.de/research/tools/likwid/}
\BIBentrySTDinterwordspacing

\bibitem{HPCG}
P.~L. Jack~Dongarra, Michael A~Heroux, ``High-performance conjugate-gradient
  benchmark: A new metric for ranking high-performance computing systems,''
  \emph{The International Journal of High Performance Computing Applications},
  vol.~30, pp. 3--10, 2015.

\end{thebibliography}
%


\end{document}